# A Simple Energy-Dependent Model for GRB Pulses with Interesting Physical Implications

Robert J. Nemiroff[a]

[a]Michigan Technological University, Department of Physics, 1400 Townsend Drive, Houghton, MI  4993

**Abstract.** A simple mathematical model for GRB pulses is postulated in both time and energy. The model breaks GRB pulses up into component functions, one general light curve function exclusively in the time dimension and four component functions exclusively in the energy dimension. Each component function of energy is effectively orthogonal to the other energy-component functions. The model is a good statistical fit to several of the most fluent separable GRB pulses known. Even without theoretical interpretation, the model may be immediately useful for fitting prompt emission from GRB pulses across energy channels with a minimal number of free parameters, sometimes far fewer than freshly fitting a GRB pulse in every energy band separately. Some theoretical implications of the model might be particularly interesting, however, as the temporal component (e.g. the shape of the light curve) is well characterized mathematically by the well known Planck distribution.

**Keywords:** gamma rays: bursts
**PACS:** 98.62.Nx, 98.70.Rz

## GRB PULSES AS A FUNCTION OF TIME

Gamma Ray Burst (GRB) prompt emission appears typically to be composed of distinct emission episodes known as "pulses" [1,2]. Although pulses usually overlap in time and energy, a fraction of GRBs feature a pulse separable enough to be analyzed by itself (see, for example, Refs. 3 and 4).  Previously, several authors have suggested relatively simple analytic forms to describe the light curves for GRB pulses. Three prominent published pulse forms in time include:

$$P(t) = A e^{-(|t-t_{max}|/\tau_{r,d})^\nu} \quad \text{(Norris et al. 1996)}[2]$$

$$P(t) = \frac{A}{(1 + t/\tau)^n} \quad \text{(Ryde et al. 2002)}[5]$$

$$P(t) = A e^{-(\tau_r/t + t/\tau_d)} \quad \text{(Norris et al. 2005)}[4]$$

The first functional form, published by Norris et al. in 1996, was fit to multiple GRB pulses detected by the Burst and Transient Source Experiment (BATSE) instrument on the Compton Gamma Ray Observatory.  Here $P(t)$ is the photon count rate as a function of time $t$, $t_{max}$ refers to the time of maximum counts, $\tau$ scales the duration of the pulse rise $r$ and decay $d$, $\upsilon$ is a peakedness parameter, and $A$ refers to the pulse amplitude.  This pulse form describes both an exponential rise and decay.

The second form shown above was used by Ryde et al. in 2002 (Ref. 5) to fit the light curves of several GRB pulses.  This form contains a power-law rise and decay. Schaefer and Dyson showed in 1996 (Ref. 6) that 10 GRB pulses detected early in BATSE's mission are only marginally well fit to an exponential decay, while a power-law decay sometimes fit better.

More recently, Norris et al. in 2005 (Ref. 4) pioneered the third functional form shown above, using fewer free parameters than the previous Norris et al. form, to fit several other isolated pulses in BATSE GRBs. Like the previous pulse form published in Norris et al. in 1996 (Ref. 2), the pulse form published by Norris et al. in 2005 (Ref. 4) also parameterizes GRB pulses with an exponential rise and an exponential decay.

All of the above pulse fitting schemes attempt to fit a pulse only at a single energy or in a single energy band. Fits to the same pulse at another energy are typically started fresh, with all the free parameters again being determined from scratch. To date, no system uses information from a pulse at one energy to fit the same pulse at another energy. Deconvolving complicated GRBs into pulses, however, can be a computationally expensive procedure.[7] Therefore, in the pulse fitting scheme published by Norris et al. in 2005 (Ref. 4), for example, the computer time taken to fit a pulse with 4 free parameters (including $t_o$, the pulse start time), in N different energy bands is 4N. The problem is not just an inefficient use of computer time -- it affects fitting accuracy as well -- information gained from fitting the pulse in a bright energy channel can be transferred to a dim energy channel.

## GRB PULSES AS A FUNCTION OF TIME AND ENERGY

The Pulse Start Conjecture and Pulse Scale Conjecture, first described by Nemiroff in 2000 (Ref. 3), posit correlations between the light curves of pulses at different energies. The Pulse Start Conjecture hypothesizes that a pulse starts at the same time $t_o$ at every energy. The Pulse Scale Conjecture posits that a pulse light curve has the same fundamental shape at all energies, for example indicated by light curve asymmetry. Fits to four of the most fluent isolated BATSE pulses ever found have shown that both the Pulse Start Conjecture and Pulse Scale Conjecture are at least approximately true for these pulses.[3] A subsequent pervasive informal investigation has shown that it is likely true for a large class of GRB pulses over the BATSE energy range, and might be true for all of them when secondary pulses are included.

These conjectures allow for mathematical generalizations of pulse light curves between energies. As a first step, it is useful to rewrite pulse light curves in terms that are independently scalable along the time and brightness axes:

$$P(t) = Ae^{-\beta(t/\tau + \tau/t)}$$ Norris Scalable

$$P(t) = \frac{A(\tau/t)^\beta}{e^{\tau/t} - 1}$$ Planck Scalable

Shown above are two light curve forms that have the requisite scalability. The first, labeled "Norris Scalable", is mathematically identical to the pulse form published by Norris et al. in 2005 (Ref. 4), but rewritten so that $\tau$ independently scales the time axis, while $A$ independently scales the brightness axis. Here $\beta$ solely determines the light curve shape -- light curves with the same $\beta$ can be exactly matched to each other by just scaling time and brightness. Therefore these scalable parameters $A$, $\tau$, $\beta$, and the implicit pulse start time $t_o$ are effectively orthogonal to each other, although renormalization might be required.

The second light curve shape, labeled "Planck Scalable" is a newly discovered form found while trying to understand what classes of functions are capable of describing GRB pulse light curves. Preliminary investigations have shown that the Planck Scalable form rivals and frequently surpasses the Norris Scalable form for best fits to the most fluent isolated BATSE pulses. This is of interest partly because Planckian functions are relatively well understood with known relations between descriptive variables. As with the Norris Scalable form, the Planck Scalable form is written so that $\tau$ independently scales the time axis, $A$ independently scales the brightness axis, and $\beta$ solely determines light curve shape. The Planck Scalable form evolves from an exponential rise to a power law decay.

In general, the free parameters $\tau$, $A$, $\beta$, and start time $t_o$ could all be functions of energy. Written in an explicit energy dependent form, the above scalable temporal pulse forms become

$$P(t, E) = A(E)e^{-\beta(E)[\frac{(t-t_o(E))}{\tau(E)} + \frac{\tau(E)}{(t-t_o(E))}]}$$  Norris Scalable (E)

$$P(t, E) = \frac{A(E)(\tau(E)/(t - t_o(E)))^{\beta(E)}}{e^{(t-t_o(E))/\tau(E)} - 1}$$  Planck Scalable (E)

In these cases, functions $A(E)$, $\tau(E)$, $\beta(E)$, and $t_o(E)$ are effectively orthogonal to each other, although as before renormalization may be required. Given these functional forms, it is possible to interpret the Pulse Start Conjecture and Pulse Scale Conjecture in terms of stated parameters. Specifically, when interpreted mathematically in terms of scalable light curve pulse shapes, the Pulse Start Conjecture posits that pulse start time $t_o$ is independent of energy, so that $t_o(E) = t_o$, a hypothesis that has been bolstered by recent analyses[8]. Similarly, the mathematical statement of the (distinct) Pulse Scale Conjecture is that $\beta$ is independent of energy: $\beta(E) = \beta$.

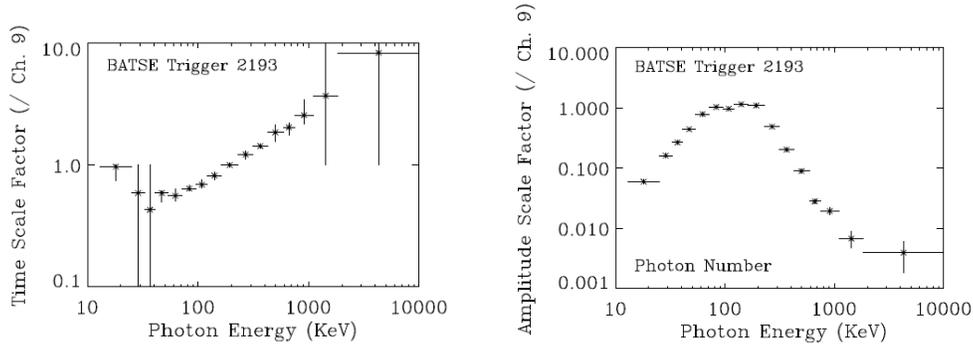

**FIGURE 1.** (a) On the left is a plot of temporal scale factor versus photon energy for the single pulse that dominated the detected flux for BATSE trigger 2193. The time stretch factor is consistent with a power law toward higher energies. The lowest energy bin might be artificially inflated due to scatter from higher energies. (b) On the right is a plot of amplitude scale factor versus photon energy for the same GRB pulse. This plot is a type of "full pulse" spectra that incorporates counts for the entire pulse in a coherent method without averaging over spectra that appear at different times.

Assuming the validity of the Pulse Start and Pulse Scale conjectures, the relative time scaling factors, $\tau(E)$, and the amplitude stretching factors, $A(E)$, can be computed for a single pulse independently across all energy bands available. Each stretch factor $\tau$ is somewhat related to the burst lag[4] factor. The above plots[3] show the widest energy range of stretch factors yet computed, though, for the fluent pulse dominating BATSE trigger 2193. The plot on the above left shows that the pulse time scale factors are highly correlated with energy, appearing consistent with a power law through the top of the BATSE energy range, so that $\tau(E) \sim E^\mu$. This power law may be useful as a cosmological invariant.

The plot on the above right shows how the amplitude scale factor $A$ varies with energy. This plot is equivalent to a full pulse spectrum. It is not a spectrum computed at any specific time in the pulse, nor a spectrum computed by integrating over the entire pulse. Rather, it involves the entire pulse by finding the best fit amplitude scaling relation between the light curve of the pulse at any two energies. Although BATSE MER channel 9 was used as the scaling base above, the shape of the full pulse spectrum in invariant to the energy of the scaling base. Note that this spectrum also does not depend on which part of the pulse is being considered, so long as enough data is available to determine an accurate scaling of the brightness of a pulse at one energy to the pulse brightness at another energy. Full pulse spectra like this may also be insightful in considering the physics of GRB pulses.

Note that the above scalable pulse forms describe pulses at a single energy $E$ only. Since GRB detectors typically measure GRB photons over a range of energies, one must be cautious about a strict interpretation of statistical fit across the duration of a pulse: a finite energy bandwidth might significantly affect the measured temporal form. Given typical hard to soft energy evolution in pulses, the beginning of a pulse may be better described by a fit at higher energy than the end of the same pulse. Since pulses typically take longer at lower energies (e.g. $\tau(E)$ decreases with $E$), a pulse with a rise best fit to data at given $E$, for example, may appear to fall below the data at the decay of the pulse where a larger $\tau$ likely applies, even if the above mathematical forms describe pulses perfectly.

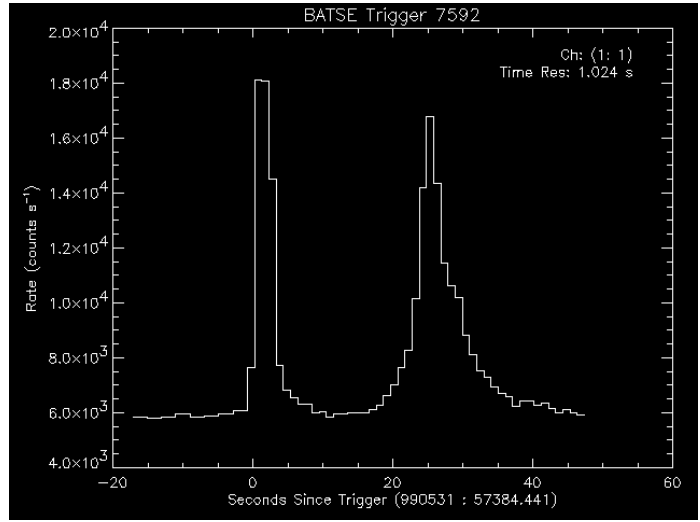

**Figure 2**. The light curve for BATSE trigger 7592 for the lowest BATSE energy channel 1, from ~20 - 50 keV.

The second pulse in Figure 2 above shows a BATSE pulse that is quite far from a good fit to either the Norris Scalable and the Planck Scalable mathematical pulse model for any parameters. The reason is currently unknown, but may be the result of the superposition of several temporally unresolved pulses, or a different physical regime operating. Informally, a few other pulses have been flagged that appear to be similarly unfitable. The plot is shown to illustrate that the above mathematical treatment has not been found to formally describe all GRB emission features that might be identified as pulses.

## ACKNOWLEDGMENTS

The author acknowledges useful conversations with Jerry Bonnell, Jay Norris, Thulsi Wickramasinghe and Amir Shahmoradi.